\documentstyle[12pt,epsfig]{article} 
\begin{document} 
\bibliographystyle{unsrt} 
 \begin{center} 
 {\large\bf
      Non-Gaussian Fluctuations in Biased Resistor Networks: 
       Size Effects versus Universal Behavior}

\vspace{0.2cm} 

C. Pennetta$^1$,  E.Alfinito$^1$, L. Reggiani$^1$, S. Ruffo$^2$ 

\vspace{0.2cm} 
$^1$ {\it Dipartimento di Ingegneria dell'Innovazione,  
Universit\`a di Lecce and \\National Nanotechnology Laboratory-INFM,  
Via Arnesano, 73100 Lecce, Italy.} 
 
\vspace{0.2cm} 
$^2$ {\it CSDC, INFN and Dipartimento di Energetica ``Sergio Stecco'', 
  \\Universit\`a di Firenze, Via S. Marta, 3, Firenze, 50139, Italy. } 

\end{center}

\vspace{0.2cm} 

{\small{\bf keywords}: Non-Gaussian Fluctuations, Resistor networks, 
Nonequilibrium stationary states, \par Disordered materials}
 

\begin{abstract}
We study the distribution of the resistance fluctuations of biased resistor 
networks in nonequilibrium steady states. The stationary conditions arise 
from the competition between two stochastic and biased processes of 
breaking and recovery of the elementary resistors. The fluctuations of 
the network resistance are calculated by Monte Carlo simulations which are 
performed for different values of the applied current, for networks of 
different size and shape and by considering different levels of intrinsic 
disorder. The distribution of the resistance fluctuations generally 
exhibits relevant deviations from Gaussianity, in particular when the 
current approaches the threshold of electrical breakdown. 
For two-dimensional systems we have shown that this non-Gaussianity is in 
general related to finite size effects, thus it vanishes in the thermodynamic 
limit, with the remarkable exception of highly disordered networks. 
For these systems, close to the critical point of the conductor-insulator 
transition, non-Gaussianity persists in the large size limit and it is 
well described by the universal Bramwell-Holdsworth-Pinton distribution. 
In particular, here we analyze the role of the shape of the network on the 
distribution of the resistance fluctuations. Precisely, we consider 
quasi-one-dimensional networks elongated along the direction of the applied 
current or trasversal to it. A significant anisotropy is found for the 
properties of the distribution. These results apply to conducting thin films 
or wires with granular structure stressed by high current densities. 
\end{abstract}


\section{Introduction and Model}
\vspace*{-0.2cm}
Strongly correlated systems usually exhibit non-Gaussian distributions of
the fluctuations of global quantities, as a consequence of the violation
of the validity conditions of the central-limit theorem. Since correlations 
become important near the critical points of phase transitions, non-Gaussian
fluctuations are usually observed near criticality 
\cite{weissman,bramwell_nat,bramwell_prl,bramwell_pre,aji,racz,jensen,antal,
gyorgyi}. 
In these conditions, the self-similarity of the system over all the
scales, from a characteristic microscopic length up to the size of the system,
has important implications on the fluctuation distribution
\cite{bramwell_nat,bramwell_prl,bramwell_pre,aji,racz,jensen,antal,gyorgyi}. 
On the other hand, far from criticality, the correlations among different
elements of the systems can also be important. This is particularly true
for systems in non-equilibrium stationary states, where non-Gaussian
fluctuations are frequently present
\cite{weissman,ausloos,derrida,raychaudhuri,pen_hcis,physicaa,spie_NG}. 
Therefore the study of non-Gaussian fluctuations and of their link with other 
features of the system can provide new insights into basic properties of 
complex systems
\cite{bramwell_nat,bramwell_prl,bramwell_pre,aji,racz,jensen,antal,gyorgyi,
derrida}.
On this respect, the observation made few years ago by Bramwell, Holdsworth 
and Pinton (BHP) \cite{bramwell_nat} of a common behavior of the distribution 
of the fluctuations in two very different systems, (the power-consumption 
fluctuations in confined turbulent-flow experiments and the magnetization 
fluctuations in the two-dimensional $XY$ model in the spin-wave regime at low 
temperature \cite{bramwell_nat,bramwell_prl,bramwell_pre}), has given rise to 
several intriguing questions about the origin of this common behavior, 
stimulating many other experimental, analytical and numerical studies. 
Successive findings have highlighted that many scale invariant systems
display the same functional form for the distribution of the fluctuations
\cite{bramwell_prl,jensen,aji,pen_hcis,physicaa,spie_NG}. Very recently,
new light on these puzzling observations has been given by Clusel et al. 
\cite{clusel}. These authors, on the basis of a study of the fluctuation
properties of the 2D $XY$ model, have proposed a criterion for 
universality-class-independent critical fluctuations \cite{clusel}. 
Actually, in the relatively simple case of the 2D $XY$ model it is possible a 
complete understanding of fluctuation phenomena. This is not possible for
nonequilibrium systems due to the lack of microscopic theories. Thus,
for these systems, we can rely only on phenomenological observations and
on analogies with better understood systems.

Here, we study the distribution of the resistance fluctuations of biased
resistor networks in nonequilibrium stationary states \cite{pen_prb04}. 
Networks of different size and shape and with different levels of internal 
disorder are considered. The resistance fluctuations are calculated by Monte 
Carlo simulations for currents close to the threshold for electrical 
breakdown. This last phenomenon consists of an irreversible increase of the 
resistance, occurring in conducting materials stressed by high current 
densities and it is associated with a conductor-insulator transition 
\cite{andersen,stan_zap,bardhan,prl_fail,bardhan1,upon02}. 
In our study we make use of the Stationary and Biased Resistor Network (SBRN) 
model \cite{SBRN_1,SBRN_2}. This model provides a good 
description of many features associated with the electrical instability of 
composite materials \cite{bardhan,bardhan1,SBRN_1} and with the 
electromigration damage of metallic lines \cite{pen_prb04,prl_fail}, 
two important classes of breakdown phenomena. 
%

We describe a thin conducting film with granular structure of length $L$, 
width $W$ and thickness $t_h \ll W, L$ as a 2D resistor network of rectangular
shape and square-lattice structure \cite{pen_prb04}. The network of resistance
$R$ is made by $N_{L}$ and $N_{W}$ resistors in the length and width 
directions respectively. Thus, the total number of resistors in the network
(excluding the contacts) is: $N_{tot}= 2N_{L}N_{W} + N_{L} -N_{W}$.
The external bias (here a constant current $I$), is applied to the network
through electrical contacts realized by perfectly conducting bars at the left 
and right hand sides of the network. The network lies on an insulating 
substrate at temperature $T_0$, acting as a thermal bath. Each resistor
has two allowed states \cite{prl_fail,gingl}: (i) regular, corresponding
to a resistance $r_{reg,n}(T_n)=r_{ref}[ 1 + \alpha (T_n - T_{ref})]$ 
and (ii) broken, corresponding to a resistance 
$r_{OP} = 10^9 r_{reg,n}(T_0) \equiv  10^9 r_0$ 
(resistors in this state will be called defects). In the above expression
$\alpha$ is the temperature coefficient of the resistance (TCR), $r_{ref}$ and
$T_{ref}$ are the reference values for the TCR and $T_n$ is the local 
temperature. The existence of temperature gradients due to current crowding 
and Joule heating effects is accounted for by taking the local temperature of 
the {\em n}-th resistor given by the following expression 
\cite{prl_fail}:

\vspace*{-0.4cm}
\begin{equation}
T_{n}=T_{0} + A [ r_{n} i_{n}^{2} + (3/4N_{neig}) 
\sum_{l=1}^{N_{neig}} ( r_{l} i_{l}^2   - r_n i_n^2)]
\end{equation} 
\vspace*{-0.3cm}

where, $i_{n}$ is the current flowing in the n{\em th} resistor and $N_{neig}$
the number of its nearest neighbors over which the summation is performed. The
parameter $A$ represents the thermal resistance of each resistor and sets the 
importance of Joule heating effects. By taking the above expression for 
$T_n$ we are assuming an instantaneous thermalization of each resistor at the 
value $T_n$ \cite{prl_fail,gingl}. In the initial state of the network 
(no external bias) we take all the resistors identical (perfect network). 
We assume that two competing biased processes act to determine the evolution 
of the network \cite{SBRN_1,SBRN_2}. These two processes consist of 
stochastic transitions between the two possible states of each resistor and 
they occur with thermally activated probabilities \cite{gingl}: 
$W_{Dn}=\exp[-E_D/k_B T_n]$ and $W_{Rn}=\exp[- E_R/k_B T_n]$, 
characterized by the two energies, $E_D$ and $E_R$ (where $k_B$ is the
Boltzmann constant). The time evolution of the network is obtained by 
Monte Carlo simulations which update the network resistance after breaking and 
recovery processes, according to an iterative procedure described in detail 
in Ref. \cite{SBRN_1}. The sequence of successive configurations provides a 
resistance signal, $R(t)$, after an appropriate calibration of the time scale. 
Depending on the stress conditions ($I$ and $T_0$) and on the network 
parameters (size, activation energies and other parameters dependent on
the material, like $r_{ref}$, $\alpha$ and $A$), the network either reaches a 
stationary state or undergoes an irreversible electrical failure 
\cite{pen_prb04,SBRN_1,SBRN_2}. This latter possibility is associated with the 
achievement of the percolation threshold, $p_c$, for the fraction of broken 
resistors. Therefore, for a given network at a given 
temperature, a threshold current value, $I_B$, exists above which electrical 
breakdown occurs \cite{SBRN_1}. 
For values of the current below this threshold, the steady state of the 
network is characterized by fluctuations of the fraction of broken resistors, 
$\delta p$, and of the resistance, $\delta R$, around their respective average
values $<p>$ and $<R>$. In particular, we underline that in the vanishing 
current limit (random percolation) \cite{prl_stat}, the ratio 
$\lambda \equiv (E_D -E_R)/k_B T_0$ determines the average fraction of defects 
and thus the level of intrinsic disorder inside the network \cite{prl_stat}. 
In the following we analyze the results of simulations performed by 
considering networks of different size and shape stressed at room temperature,
$T_0=300$ K, by a current density $j \equiv I/N_W = 0.32$ mA. We have taken: 
$\alpha=3.6 \times 10^{-3}$ K$^{-1}$, $T_{ref}=273$ K, 
$r_{ref} = 0.048$ $\Omega$,  $A=2.7 \times 10^8$ K/W, $E_{D}=0.41$ eV and 
$E_{R}=0.35$ eV. This choice of the parameters is appropriate to describe
the behavior under electromigration of metallic lines of Al-0.5\%Cu 
studied in Ref. \cite{pen_prb04} and it corresponds to studying a network
with an intermediate level of intrinsic disorder.   
\begin{figure}
\begin{center}
   \includegraphics[height=.12\textheight]{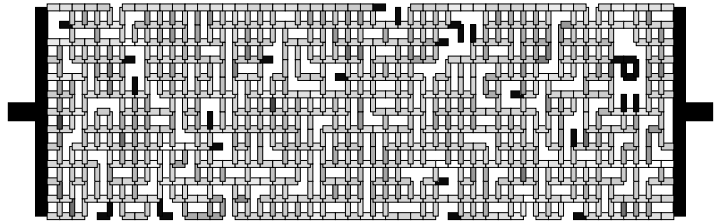}
   \caption{Pattern of a network $12 \times 50$ stressed by a current
   density $j=0.32$ mA. The grey boxes show the backbone of
   the network, the black ones the "dangling bonds" (branches
   with zero-current, while the missing boxes correspond to the broken
   resistors. This pattern has been calculated at $t=4 \times 10^4$ 
   (time expressed in iteration steps). }
\end{center} 
\end{figure} 

\begin{figure}
\begin{center}
  \includegraphics[height=.22\textheight]{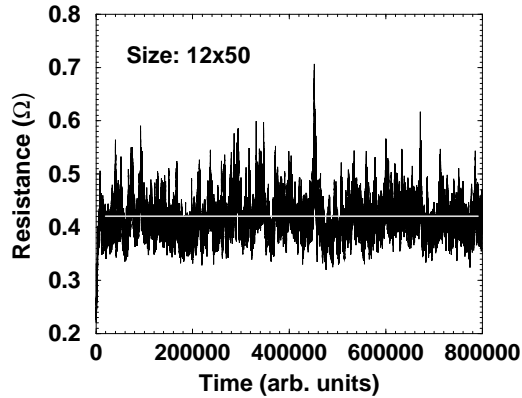}
 \caption{Resistance evolution of the network in Fig. 1. The 
   time is expressed in arbitrary units (iteration steps), the resistance
   in Ohm. The grey line shows the average value of the resistance.}
\end{center}
\end{figure} 
%


\section{RESULTS AND CONCLUSIONS}
\vspace*{-0.2cm}
Figure 1 displays the pattern of a network $12 \times 50$ calculated at a 
given time, $t = 4 \times 10^4$, (expressed in iteration steps) in the 
stationary regime of the network, i.e. for $t > \tau_{rel}$, where 
$\tau_{rel} \approx 8 \times 10^3$ is the relaxation time for the achievement 
of the nonequilibrium stationary state. The network in this figure is stressed
by a current density ($j=0.32$ mA) close to the breakdown value, 
$j_B=I_B/N_W$. The resistance evolution for the same network is reported in 
Fig. 2. In this figure the grey line shows the average value of the 
resistance, $<R>$. We note that both the average resistance and the relative 
variance of the resistance fluctuations, $<(\delta R)^2>/<R>^2$, depend on 
$j$. A detailed analysis of the behavior of these two quantities as a function
of the current can be found in Refs. \cite{SBRN_1,SBRN_2}.
In previous works \cite{pen_hcis,physicaa,spie_NG} we have analyzed the 
effects on the distribution of the resistance fluctuations of the biasing
current \cite{pen_hcis}, of the intrinsic disorder and of the size of the 
network \cite{physicaa,spie_NG}, by limiting ourself to discuss square 
networks. Here, we focus our discussion on shape effects: precisely we analyze
the effect on the distribution of the fluctuations of scaling the size of 
the network separately in the two directions, i.e. of scaling  separately 
the width, $N_W$ and the length, $N_L$, of the network.
Figure 3(a) shows the distributions of the resistance fluctuations obtained 
for two networks of size $12 \times 50$ (big circles) and $50 \times 12$ 
(triangles) stressed by the same current density. In this figure (and in the 
followings) we denote with $\Phi$ the probability density function (PDF) of 
the distribution and with $\sigma$ the root mean square deviation from the 
average value. This normalized representation, by making the distribution 
independent of its first and second moments, is particularly convenient to 
explore the functional form of a distribution \cite{bramwell_prl}. 
A lin-log scale is adopted for convenience to plot the product $\sigma \Phi$ 
as a function of $(<R>-R)/\sigma$. The PDFs in Fig. 3 and all the others in 
this paper have been calculated by considering time series containing about 
$10^6$ resistance values. For comparison, in Fig. 3 we also report the 
Gaussian distribution (dashed curve) and the BHP distribution (continuous 
curve) \cite{bramwell_nat,bramwell_prl}. The PDF obtained for the network 
$12 \times 50$ (corresponding to the signal in Fig. 2) exhibits a strong 
non-Gaussianity, well described by the BHP curve. 
By contrast, the PDF obtained for the network $50 \times 12$ is nearly 
Gaussian. At a first insight, this result can seem surprising: in fact the 
two networks are composed by nearly the same number of resistors, moreover 
the dissipated electric power per unit volume, $RI^2/(LW) \propto j^2$, is the
same in both cases. As a consequence, the average fraction of defects 
$p \approx 0.19$ is also the same. However, the percolation threshold $p_C$ is
different for the two networks \cite{pen_prb04}. Therefore, the nearly 
Gaussian distribution of the $50 \times12$ network is due to the higher value 
of $p_C$ (and thus to the higher stability) of this network \cite{pen_prb04}. 
For comparison, we report in Fig. 3(b) the PDFs calculated for two square 
networks $12 \times 12$ and $50 \times 50$ biased by the same current density. 
Again, the dissipated power density and the average fraction of defects are 
the same for both networks. However, for square $N \times N$ networks the
percolation threshold is roughly independent of the size, even for biased
percolation \cite{pen_prb04}. Thus, for these networks the higher instability 
and the stronger non-Gaussianity for decreasing $N$ is mainly related with the 
increase in magnitude of the fluctuations associated with the smaller size 
\cite{pen_hcis,physicaa}.

The normalized PDFs of the resistance fluctuations calculated for
several networks of different size are reported in Fig. 4. Figure 4(a)
displays PDFs obtained for networks elongated along the direction of the 
applied current (precisely networks of a given width, $N_W=12$, and with
increasing length, $N_L = 50\div 400$), while Fig. 4(b) shows PDFs obtained 
for networks elongated in a direction trasversal to the current applied
(precisely networks of a given length, $N_L=12$, and with increasing width, 
$N_W = 50\div 400$. We can see that for trasversal networks the PDF is rather 
insensitive to the width and the small non-Gaussianity for small widths 
completely vanishes already for networks with $N_W=200$. By contrast, for
longitudinal networks, the PDF is sensitive to the length. However, it should
be noted that the PDF obtained for $N_L=200$ practically overlaps
with that obtained for $N_L=400$ and both exhibit non-Gaussian tails.
Since the correlation length, $\xi$, for these networks is estimated to be
$\xi < 5$, networks with $N_L=400$ can be considered as infinitely long.
Thus, Fig. 4(a) suggests a persistent non-Gaussianity for longitudinal
networks in the limit $N_L \rightarrow \infty$, associated with the finite
size of the network in the transversal direction. Furthermore, the
magnitude of this non-Gaussianity is expeceted to be controlled by the
level of intrinsic disorder. 

In conclusions, we have studied the distribution of the resistance
fluctuations of biased resistor networks in nonequilibrium stationary
states. We have considered networks biased by currents close to the threshold 
of electrical breakdown. As a general trend, the distribution of the
fluctuations is found to exhibit relevant deviations from Gaussianity,
which are in general related to finite size effects \cite{pen_hcis,physicaa}. 
However, for systems close to the critical point of the conductor-insulator
transition, the non-Gaussianity persists in the large size limit 
\cite{pen_hcis,physicaa} and it is well described by the universal
Bramwell-Holdsworth-Pinton distribution. Furthermore, we have analyzed the
role of the shape of the network on the distribution of the resistance 
fluctuations, by considering quasi-one-dimensional networks elongated
along the direction of the applied current or trasversal to it. 
A significant anisotropy is found for the properties of the distribution.
These results apply to conducting thin films or wires with granular
structure stressed by high currents.

\begin{figure}
\begin{center}
\includegraphics[height=.20\textheight]{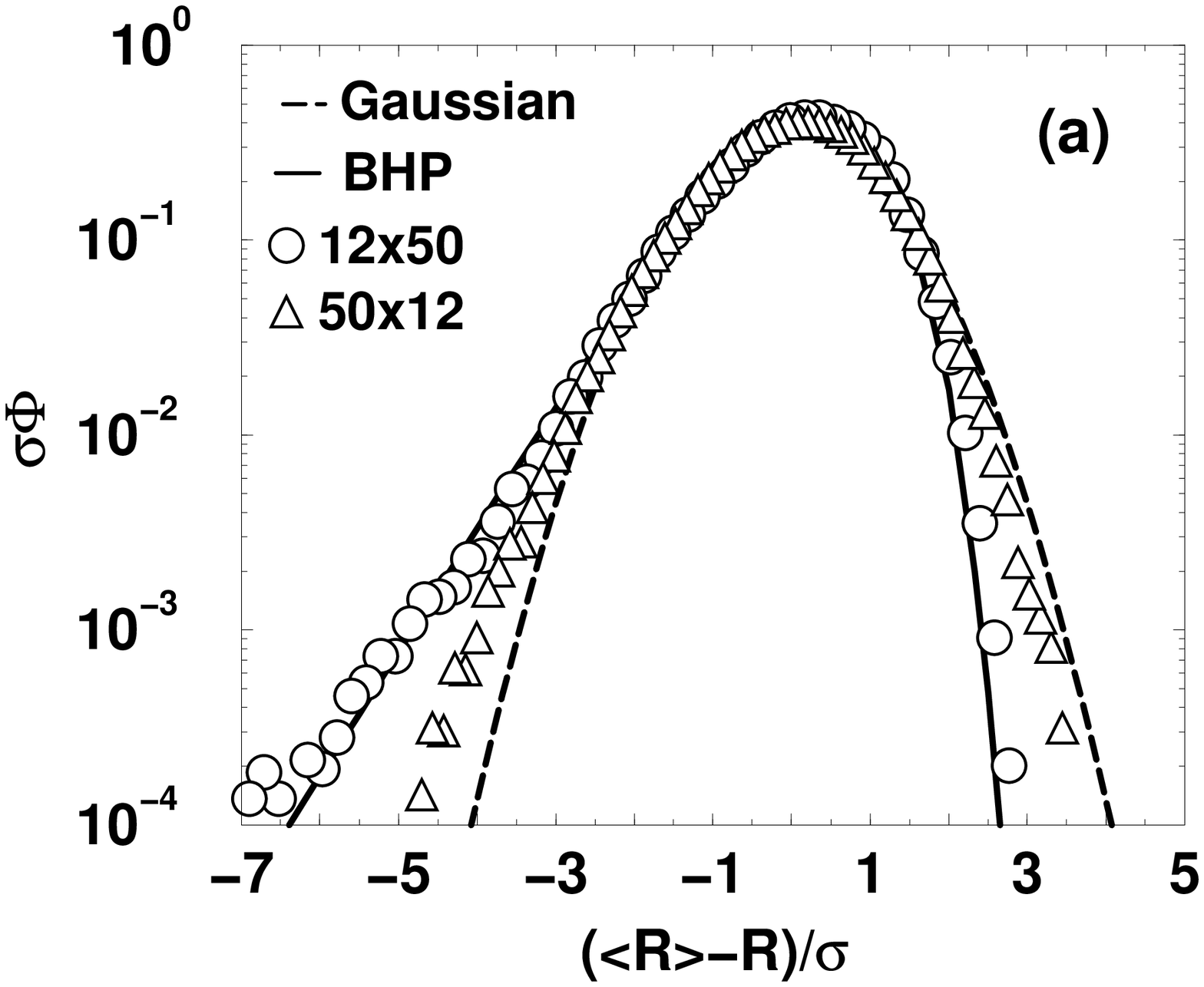}
\hspace*{0.5cm}
\includegraphics[height=.20\textheight]{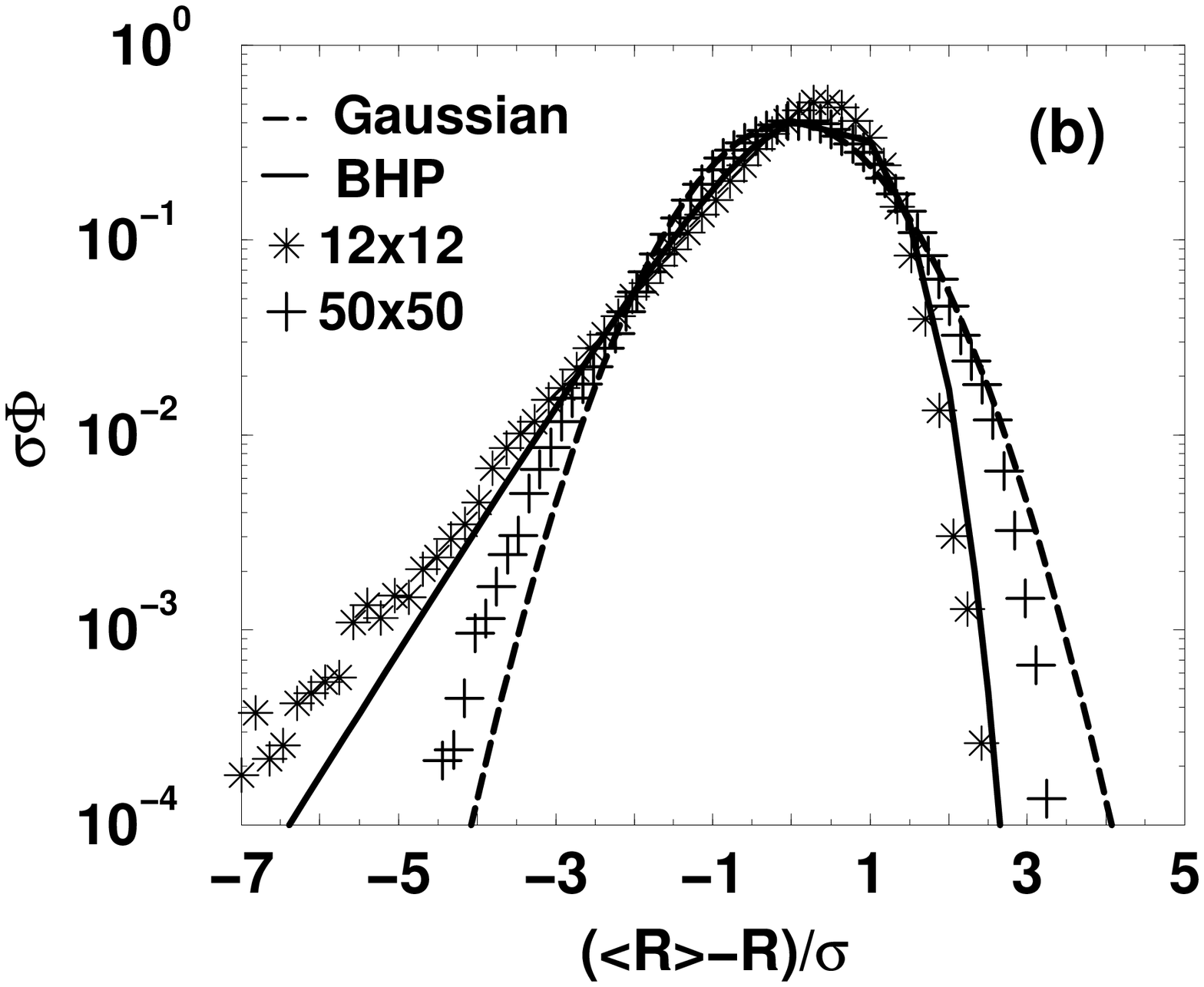}
   \caption{Normalized PDF of the resistance fluctuations of networks of
   different size and stressed by the same current density 
   $j=0.32$ mA. Precisely, in (a) the size is: $12 \times 50$ (big circles) 
   and $50 \times 12$ (triangles); in (b): $12\times 12$ (stars), 
   $50 \times 50$ (plus). The solid and dashed curves 
   refer to the BHP and Gaussian distributions, respectively.}
\end{center}
\end{figure} 

\begin{figure}
\begin{center}

 \includegraphics[height=.20\textheight]{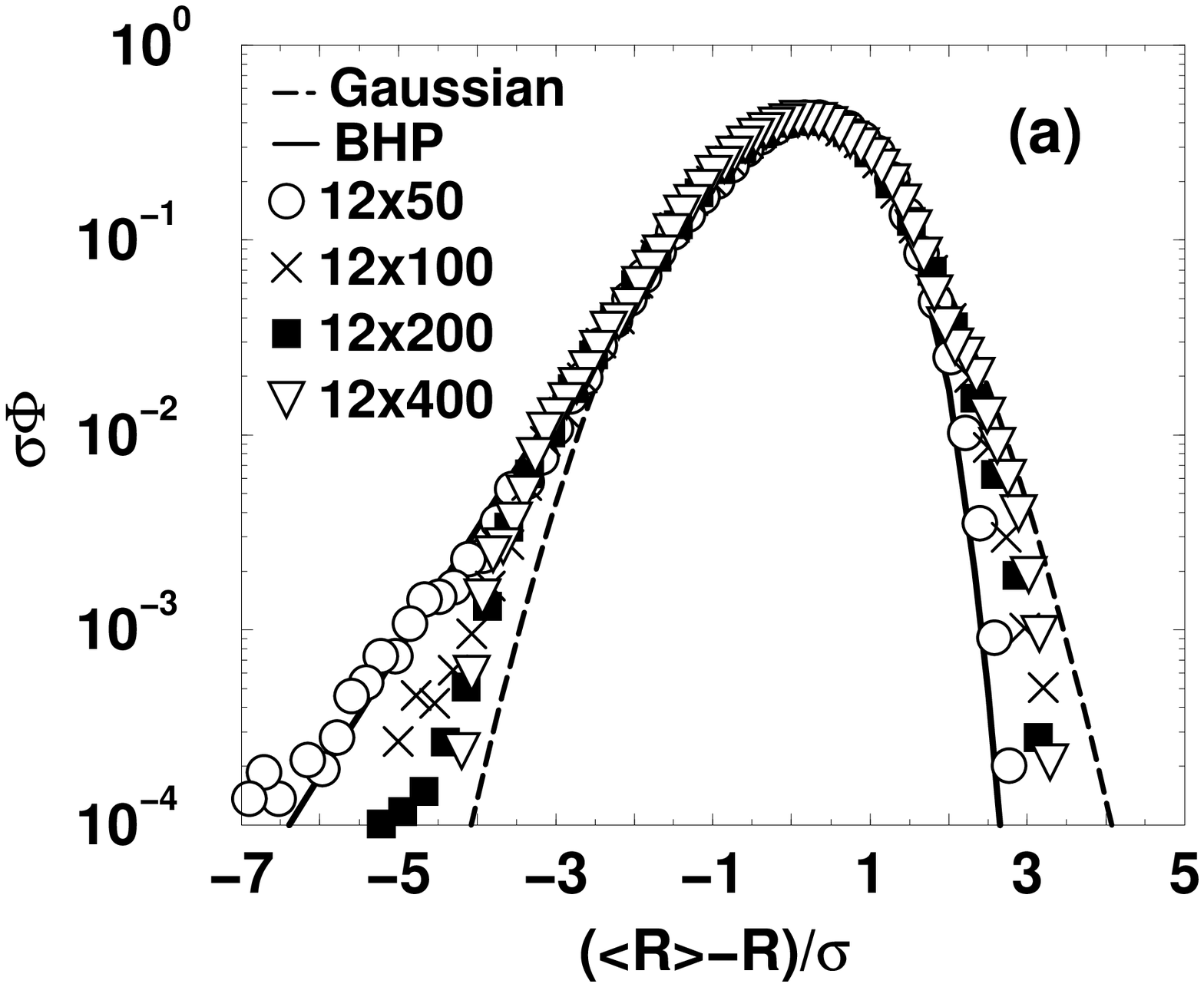}
\hspace*{0.5cm}
\includegraphics[height=.20\textheight]{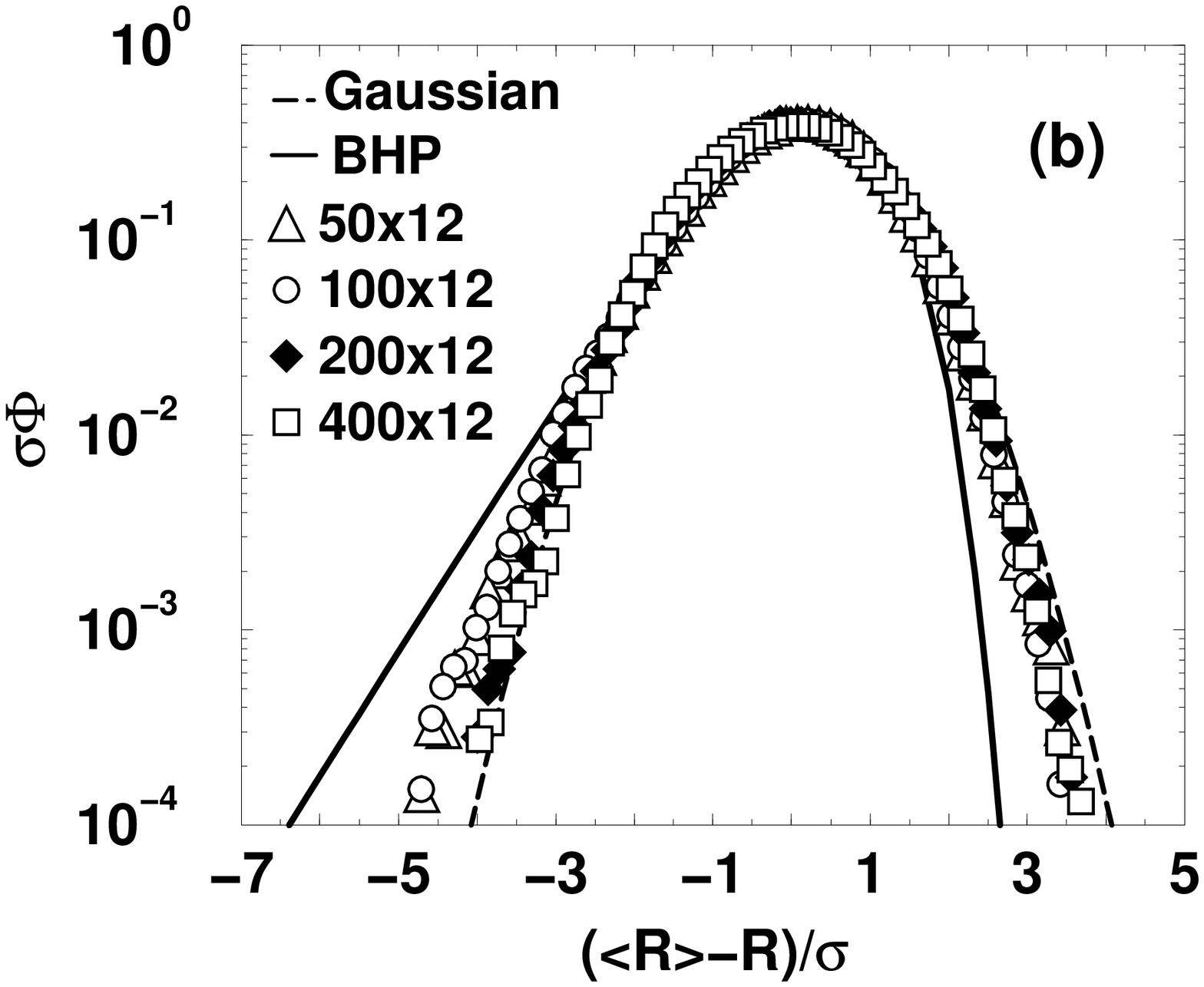}
  \caption{Normalized PDF of the resistance fluctuations of networks of
   different size and stressed by the same current density 
   $j=0.32$ mA. Precisely, in (a) the size is:
   $12 \times 50$ (big circles), $12 \times 100$ (crosses), 
   $12 \times 200$ (full squares) and $12 \times 400$ (down triangles); 
   in (b) the size is: $50 \times 12$ (triangles), $100 \times 12$ 
   (small circles), $200 \times 12$ (full diamonds) and 
   $400 \times 12$ (squares). The solid and dashed 
   curves have the same meaning of Fig. 3.}
 \end{center}
\end{figure}

\vspace{0.2cm}

{\large\bf Acknowledgments}

\vspace*{0.2cm}

This work has been performed within the cofin-03 project ``Modelli e misure 
di rumore in nanostrutture'' financed by Italian MIUR, the SPOT NOSED project 
IST-2001-38899 of EC is also acknowledged. P.C.W. Holdsworth and S. Caracciolo
are gratefully acknowledged for helpful discussions.


\bibliography{nongau}

\end{document}